\documentclass{optica-article}
\usepackage{braket}
\journal{oe}

\articletype{Research Article}

\usepackage{lineno}
\linenumbers

\begin{document}
\nolinenumbers

\title{Frequency-domain engineering of bright squeezed vacuum for continuous-variable quantum information}

\author{Inbar Hurvitz,\authormark{1,*} Aviv Karnieli,\authormark{2} and Ady Arie\authormark{1}}

\address{\authormark{1}School of Electrical Engineering, Fleischman Faculty of Engineering, Tel Aviv University, Tel Aviv 69978, Israel\\
\authormark{2}Sackler School of Physics, Faculty of Exact Sciences, Tel Aviv University, Tel Aviv 69978, Israel}

\email{\authormark{*}inbarhurvitz@mail.tau.ac.il} 



\begin{abstract*}
Multimode bright squeezed vacuum is a non-classical state of light hosting a macroscopic photon number while offering promising capacity for encoding quantum information in its spectral degree of freedom. Here, we employ an accurate model for parametric down-conversion in the high-gain regime and use nonlinear holography to design quantum correlations of bright squeezed vacuum in the frequency domain. We propose the design of quantum correlations over two-dimensional lattice geometries that are all-optically controlled, paving the way toward continuous-variable cluster state generation on an ultrafast timescale. Specifically, we investigate the generation of a square cluster state in the frequency domain and calculate its covariance matrix and the quantum nullifier uncertainties, that exhibit squeezing below the vacuum noise level. 
\end{abstract*}

\section{Introduction}
The spectral degree of freedom of photons has become attractive thanks to its potentially unlimited Hilbert space dimension and to its large bandwidth. These are appealing traits for quantum information protocols  \cite{Clemmen2016,Karnieli2018, Shaked2018}, that were also shown useful, for example, in quantum key distribution \cite{Bloch2007} and quantum sensing \cite{Clemmen2016}. The advent of quantum optical frequency combs \cite{Kues2019} allowed the generation and manipulation of high-dimensional photonic qudits \cite{Lu2022, Kues2017,entanglement_of_60_modes_Pfister}, and even the formation of frequency-time bin cluster states \cite{Reimer2019,Chekhova_Science_2022}. 
In particular, shaping the spectrum of photon pairs generated via spontaneous parametric down-conversion (SPDC) by second-order nonlinearity in crystals \cite{Dosseva,Graffitti2017} and waveguides \cite{Helt2015}, third-order nonlinearity in spontaneous four-wave mixing in fibers and resonators \cite{SFWM_2023} or in hot atomic vapors \cite{Da_Zhang} has drawn tremendous attention in the past few decades. Here we focus on the first method, whereby the progress in engineered nonlinear photonic crystals (NLPC) \cite{Shiloh} allows spectral-temporal entanglement \cite{Fedrizzi2022,Graffitti:18}.

In parallel to the advances of such discrete-variable quantum optical schemes in the frequency domain, there was also progress in continuous-variable (CV) schemes employing the spectral degree of freedom. The main resource for frequency-domain CV quantum light is the multimode squeezed vacuum \cite{Sharapova2018}, demonstrating quantum correlations on the one hand, while containing a large number of spectral modes allowing high quantum information capacity \cite{Sharapova2018}, and even the realization of CV cluster states on a frequency comb \cite{Pfister_2019,entanglement_of_60_modes_Pfister}. Bright squeezed vacuum (BSV), a squeezed vacuum state containing a macroscopic number of photons, can be generated in strongly-pumped NLPCs, pushing SPDC to the high-gain regime. Tailoring the spectrum of BSV is important for several applications such as CV cluster state generation \cite{Pfister_2019}, broadband homodyne detection \cite{Shaked2018} and supersensitive phase measurements \cite{Supersensitive_Phase_Meas2017}. While nonlinear holography techniques together with pump shaping are well-established methods for engineering SPDC photons, their full potential has not yet been deployed for the frequency-domain engineering of BSV. To date, the most prevalent schemes for spectral shaping of BSV use either SU(1,1) interferometers \cite{Sharapova2018}, or employ the frequency comb of an optical parametric oscillator (OPO) \cite{Menicucci2007,Cai2017} together with spectrally-shaped pumps.

In this paper, we analyze a new scheme for frequency-domain engineering of BSV using a combination of nonlinear holograms and shaped pump spectra. Our model extends recent proposal by Drago \cite{Drago2022} to regime of high nonlinear gain. We validate our model against known results of SPDC shaping using engineered NLPCs extended to the high-gain regime, and then propose novel designs of high-dimensional frequency bin correlations with square lattice geometry. This capability allows for the generation of CV cluster states in the frequency domain, which are important for measurement-based quantum computing (MBQC) and quantum networks \cite{Treps2018,Nokkala_2018}, without the necessity of an OPO frequency comb. Using our holographic technique, we propose an alternative method to generate such states by simultaneously using nonlinear holograms and shaped pump spectra, which provide additional degrees of freedom. For example, we show how the quantum correlations can be all-optically controlled on an ultrafast timescale using shaped pump pulses interacting with the same nonlinear holograms. Finally, we analyze CV cluster states for a simple class of designs and demonstrate their expected quadrature correlations and squeezing.

\section{Theoretical model}
\begin{figure}[h]
\centering
\noindent
\includegraphics[width=0.8\columnwidth]{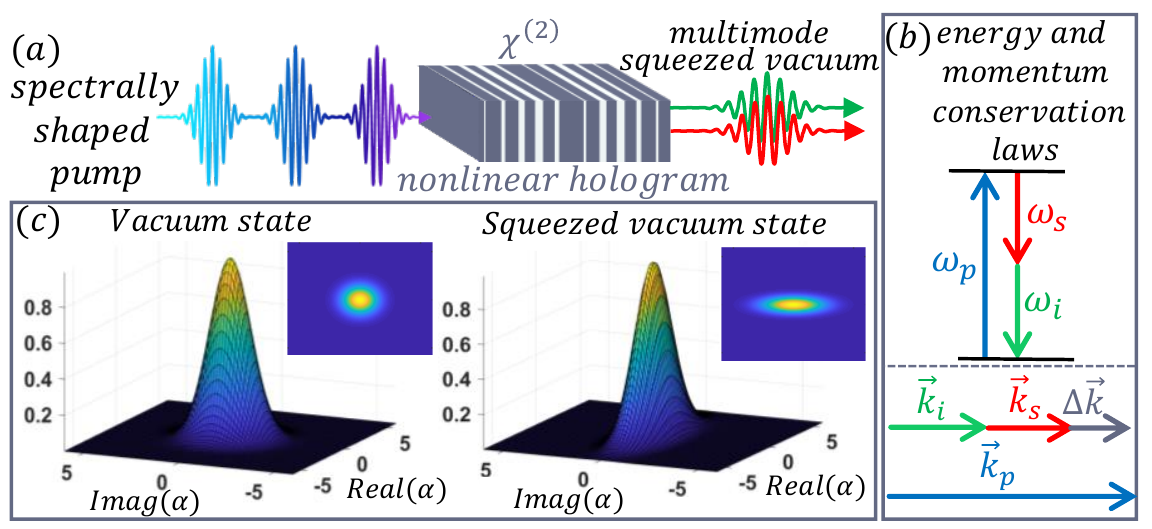}
\caption{Illustration of multimode squeezed light generation in the high gain regime. (a) An intense pump pulse with a shaped spectrum enters a modulated NLPC hosting a spectral hologram. Each photon in the pump beam (of frequency $\omega_p$) down converts to two photons (with frequencies $\omega_s$,$\omega_i$) under conservation of energy and momentum, summarized in (b). Panel (c) shows Wigner functions of a vacuum state (left) and a squeezed vacuum state (right). For Gaussian states such as squeezed vacuum, the moments of the Wigner function in phase space are all prescribed by first-order correlations of the form of Eq.\ref{eq: G1} and Eq.\ref{eq: Q_ij}.} 
\label{fig:concept}
\end{figure}

Several accurate models exist for the description of multimode squeezed light generation, for example, phase-space methods investigating the effects of ultrashort pump spectra \cite{Drummond1995} or predicting spatial correlations \cite{Brambilla}. More recent theoretical descriptions of BSV states and their correlations were introduced \cite{Sharapova2020}, predicting dynamics, broadening of angular distribution, and Schmidt ranks as the squeezing parameter is increased.  Here, we extend these models to incorporate the combined effects of high-gain, shaped pump spectra and modulated NLPCs (see Fig.\ref{fig:concept}), while employing a simulation scheme based on a recently developed and experimentally-validated numerical simulation tool for SPDC \cite{TM2020,Rozenberg}. We focus on the case of a single spatial mode with no diffraction, for a type-II SPDC process in a $\mathrm{KTiOPO_4}$ (KTP) engineered NLPC, where the \textit{idler} ($i$) and \textit{signal} ($s$) fields differ by polarization. While this is an appropriate model for plane-wave pumps in on-axis phase-matched NLPCs, it can also be applied to nonlinear waveguides.

We begin by considering the quantum Maxwell's equations \cite{Garrison2008} for the frequency-dependent field operators \cite{Rozenberg}, under the undepleted pump and slowly-varying envelope approximations,  
\begin{equation}
\begin{split}
i\frac{\partial}{\partial z} \left(\begin{matrix}\hat{A}_i\\\hat{A}_s\end{matrix}\right)&(z,\omega)  = -\frac{\omega^2\chi^{(2)}\mathcal{E}_p}{c^2 k(\omega)}d_{\mathrm{NL}}(z) \int d\omega' \mathrm{e}^{i\Delta k(\omega,\omega')z}\mathcal{P}(\omega+\omega')\left(\begin{matrix}\hat{A}_s^{\dagger}\\\hat{A}_i^{\dagger}\end{matrix}\right)(z,\omega'),
\end{split}
\label{eq:QuantumMaxwell}
\end{equation}
where $\hat{A}_i,\hat{A}_s$ are, respectively, the idler and signal field slowly-varying-envelope operators \cite{Garrison2008}, $z$ is the propagation coordinate, $\chi^{(2)}$ is the value of the quadratic nonlinear coefficient, $d_{\mathrm{NL}}(z)$ is the modulation of the NLPC, $\mathcal{E}_p$ is the pump electric field, $\mathcal{P}(\omega)$ is the normalized pump spectrum, $c$ is the speed of light, $k(\omega)$ is the frequency-dependent wavenumber of the medium, and $\Delta k= k_p(\omega+\omega')-k_i(\omega)-k_s(\omega')$ is the phase mismatch parameter.

We wish to study the quantum correlations of the generated multimode BSV in the frequency domain. It is well known that squeezed vacuum is a zero-mean Gaussian state \cite{Adesso_2014}, which could be described in phase space by a Wigner function of the form \cite{Adesso_2014}:
\begin{equation}
    W(\Vec{\alpha})=\frac{1}{\pi^N} \frac{1}{\det \mathbf{\sigma}}\exp{(-\Vec{\alpha}^\dagger\sigma^{-1}\Vec{\alpha})}
    \label{eq: Wigner}
\end{equation}
where $N$ is the number of modes, $\Vec{\alpha}=(\alpha_1,\alpha_2,...,\alpha_N,\alpha_1^*,\alpha_2^*,...,\alpha_N^*)$ where $\alpha_j$ are complex numbers, and the covariance matrix $\sigma$, of size $2N$ by $2N$, is defined by
\begin{equation}
    \sigma_{mn} = \braket{\xi_m \xi_n^{\dagger} + \xi_m^{\dagger} \xi_n}.
\label{eq:covariance}
\end{equation}
where $\vec{\xi} = (a_1,a_2,...,a_N,a_1^\dagger,a_2^\dagger,...,a_N^\dagger)$ is a vector operator of length $2N$ containing all annihilation and creation operators. Note that $\sigma_{mn}$ may contain all terms of the form $\braket{a_ma_n},\braket{a_m^\dagger a_n}$ and $\braket{a_m^\dagger a_n^\dagger}$. Eqs.\ref{eq: Wigner} and \ref{eq:covariance} imply that first-order quantum correlations (i.e., those that are quadratic in the field operators) hold all the information on the quantum state. Therefore, in the following, we consider the first-order quantum correlations of BSV, and derive all relevant observables from them.

To calculate the first-order quantum correlations, we begin by observing that the initial state is vacuum ($\ket{\psi}=\ket{0}$) for which the first-order correlation function can be written in terms of the field operators (Eq.\ref{eq:QuantumMaxwell}) propagated from $z=0$ to $z=L$ (where $L$ is the crystal's length) as
\begin{equation}
\begin{split}
G_{\alpha\beta}^{(1)} (\omega, \omega') =
\bra{0}\hat{A}_{\alpha}^{\dagger}(L,\omega)\hat{A}_{\beta}(L,\omega')\ket{0},
\label{eq: G1}
\end{split}
\end{equation}
where $\alpha,\beta=i,s$. We insert the resolution of identity in the Fock basis $I=\int d\Omega \sum_{n,\gamma} \ket{{n_{\Omega,\gamma}}}\bra{{n_{\Omega,\gamma}}}$ into Eq.\ref{eq: G1}, where $\gamma=i,s$. As the dynamics is linear and mixes creation and annihilation operators, the field operator can only couple the vacuum state to a single photon state $\ket{{1_{\Omega,\gamma}}}$. Namely, the transition amplitudes of the form $\braket{{n_{\Omega,\gamma}}|\hat{A}_{\alpha}(L,\omega)|0}$ are all zero except for $n=1$. Following Refs \cite{TM2020,Rozenberg,Sipe2014,Sipe2020,Christ2013}, we write the simplified first-order correlation function in terms of these single-photon amplitudes as
\begin{equation}
\begin{split}
G_{\alpha\beta}^{(1)} (\omega, \omega') = \int d\Omega \sum_{\gamma}\bra{0}&\hat{A}_{\alpha}^{\dagger}(L,\omega)\ket{{1_{\Omega,\gamma}}}\bra{{1_{\Omega,\gamma}}}\hat{A}_{\beta}(L,\omega')\ket{0}.
\label{eq: G1 final}
\end{split}
\end{equation}
Out of the eight possible combinations for single photon amplitudes ($\alpha,\beta,\gamma=i,s$), it is readily shown that four are identically zero for all $z$. The remaining four, non-vanishing amplitudes are defined as
\begin{equation}
\begin{split}
& A_{i,\Omega}^{\mathrm{vac}}(z,\omega) \equiv \bra{0} \hat{A}_{i}(z,\omega) \ket{{1_{\Omega,i}}} \\& A_{i,\Omega}^{\mathrm{out}}(z,\omega) \equiv \bra{{1_{\Omega,{s}}}} \hat{A}_{i}(z,\omega) \ket{0}
\\& A_{s,\Omega}^{\mathrm{vac}}(z,\omega) \equiv \bra{0} \hat{A}_{s}(z,\omega) \ket{{1_{\Omega,s}}} \\& A_{s,\Omega}^{\mathrm{out}}(z,\omega) \equiv \bra{{1_{\Omega,{i}}}} \hat{A}_{s}(z,\omega) \ket{0},
\label{eq: A_vac,A_out}
\end{split}
\end{equation}
where $i,s$ denote the idler and signal respectively. These four complex field amplitudes can be interpreted as the vacuum field ($A_{\alpha,\Omega}^{\mathrm{vac}}$) and the generated output fields ($A_{\alpha,\Omega}^{\mathrm{out}}$) in the signal and idler modes, originating from the vacuum fluctuations in frequency $\Omega$. From Eq.\ref{eq:QuantumMaxwell} we obtain their dynamics in terms of four integro-differential equations
\begin{equation}
\begin{split}
& i\frac{\partial}{\partial z} \left(\begin{matrix}A_{i,\Omega}^{\mathrm{out}}\\A_{i,\Omega}^{\mathrm{vac}}\end{matrix}\right)(z,\omega)  = -\frac{\omega^2\chi^{(2)}\mathcal{E}_p}{c^2 k(\omega)}d_{\mathrm{NL}}(z)   \int d\omega' \mathrm{e}^{i\Delta k(\omega,\omega')z}\mathcal{P}(\omega+\omega')\left(\begin{matrix}A_{{s},\Omega}^{\mathrm{vac*}}\\A_{{s},\Omega}^{\mathrm{out*}}\end{matrix}\right)(z,\omega')
\\& i\frac{\partial}{\partial z}\left(\begin{matrix}A_{s,\Omega}^{\mathrm{out}}\\A_{s,\Omega}^{\mathrm{vac}}\end{matrix}\right)(z,\omega)  = -\frac{\omega^2\chi^{(2)}\mathcal{E}_p}{c^2 k(\omega)}d_{\mathrm{NL}}(z)   \int d\omega' \mathrm{e}^{i\Delta k(\omega,\omega')z}\mathcal{P}(\omega+\omega')\left(\begin{matrix}A_{{i},\Omega}^{\mathrm{vac*}}\\A_{{i},\Omega}^{\mathrm{out*}}\end{matrix}\right)(z,\omega').
\end{split}
\label{eq:dynamics}
\end{equation}
Whereas Eq.\ref{eq:QuantumMaxwell} was for field operators, Eq.\ref{eq:dynamics} can be computed numerically by a spit step algorithm, since it describes the evolution of the four complex amplitudes described in Eq.\ref{eq: A_vac,A_out}.  From here onwards, we use the normalized gain parameter $\kappa = \omega \chi^{(2)}\mathcal{E}_p L/c n$ to characterize the high-gain regime, for which $\kappa \geq 1$. The boundary conditions at $z=0$ are given as $A_{\alpha,\Omega}^{\mathrm{out}}(0,\omega)=0$ and $A_{\alpha,\Omega}^{\mathrm{vac}}(0,\omega)=\delta(\omega-\Omega)\mathcal{E}_{0,\alpha}(\omega)$, where $\mathcal{E}_{0,\alpha}(\omega)$ is the c-number vacuum field amplitude at frequency $\omega$. When the fields at $z=L$ are found, we compute Eq.\ref{eq: G1 final} by tracing over all possible vacuum modes ${\Omega}$, 
\begin{equation}
\begin{split}
G_{\alpha\beta}^{(1)} (\omega, \omega') =\delta_{\alpha,\beta} \int d\Omega A_{\alpha,\Omega}^{\mathrm{out}*}(L,\omega)A_{\beta,\Omega}^{\mathrm{out}}(L,\omega'),
\label{eq: G1 fields}
\end{split}
\end{equation}
where the Kronecker delta in Eq.\ref{eq: G1 fields} arises due to the vanishing photon amplitudes other than those appearing in Eq.\ref{eq: A_vac,A_out}, implying that $G^{(1)}_{is}=G^{(1)}_{si}=0$.

A second type of first-order quantum correlation involves terms that are quadratic in the annihilation or creation operators, e.g. terms of the form $\braket{a_m a_n}$ or $\braket{a_m^\dagger a_n^\dagger}$. We denote such terms as
\begin{equation}
Q_{\alpha\beta} (\omega, \omega')= \bra{0}\hat{A}_{\alpha}(L,\omega)\hat{A}_{\beta}(L,\omega')\ket{0}  
\end{equation}
and, employing a similar procedure to the one used to derive the first-order correlation, Eq.\ref{eq: G1 fields}, we calculate these correlations in terms of the complex field amplitudes as
\begin{equation}
\begin{split}
Q_{\alpha\beta} (\omega, \omega') & = 
(1-\delta_{\alpha,\beta})\int d\Omega A_{\alpha,\Omega}^{\mathrm{vac}}(\omega) A_{\beta,\Omega}^{\mathrm{out}}(\omega'),
\label{eq: Q_ij}
\end{split}
\end{equation}
where the Kronecker delta appears for a similar reason as in Eq.\ref{eq: G1 fields}, implying that $Q_{ii} =Q_{ss} =0$.

Equipped with all relevant first-order correlations, we can now calculate any desired observable. The first one is the second-order correlation function, which predicts important measurables in quantum optics experiments in the low-gain regime of SPDC, which will be our base case for validation of our model. Namely, the second order correlation function $G_{\alpha\beta\beta\alpha}^{(2)}(\omega, \omega'; \omega', \omega)=\langle \hat{A}^{\dagger}_{\alpha}(z,\omega)\hat{A}^{\dagger}_{\beta}(z,\omega') \hat{A}_{\beta}(z,\omega') \hat{A}_{\alpha}(z,\omega)\rangle$ \cite{Brambilla} denotes the probability of finding one $\alpha$ photon in mode $\omega$ and another $\beta$ photon in mode $\omega'$, and corresponds to the coincidence probability of these two photons. For Gaussian quantum states, such as the multimode squeezed vacuum, the second-order correlation function can be written in terms of first-order correlations as
\begin{equation}
\begin{split}
G_{\alpha\beta\beta\alpha}^{(2)}(\omega, \omega'; \omega', \omega)  =   &      G_{\alpha\alpha}^{(1)}(\omega,\omega) G_{\beta\beta}^{(1)}(\omega',\omega')   + |G_{\alpha\beta}^{(1)}(\omega,\omega')|^2+  |Q_{\alpha\beta}(\omega,\omega')|^2.
\label{eq: G2_derivation}
\end{split}
\end{equation}
In the high-gain regime, an important observable that is commonly employed to evaluate multimode BSV is the noise reduction factor (NRF), which quantifies twin-beam squeezing \cite{Sharapova2018,two_color_BSV}. In terms of the first-order correlations we obtain from our simulation, it can be expressed as follows
\begin{equation}
\begin{split}
\mathrm{NRF}\equiv \frac{\mathrm{Var}(N_i-N_s)}{\braket{N_i+N_s}} = 1 + \frac{\mathcal{N}^2_i+\mathcal{N}^2_s-\mathcal{Q}_{is}^2-\mathcal{Q}_{si}^2}{\mathcal{N}_i+\mathcal{N}_s},
\label{eq: NRF}
\end{split}
\end{equation}
where $N_i, N_s$ are the number of photons of the idler and signal, $\mathcal{N}_j=\int d\omega G^{(1)}_{jj}(\omega,\omega)$ and $\mathcal{Q}_{jk}^2=\int d\omega \int d\omega' |Q_{jk}(\omega,\omega')|^2$ denote observables integrated over a given frequency window, equivalent to the spectral bandwidth of a filter in an experiment \cite{two_color_BSV}. For twin-beam squeezing, $\mathrm{NRF}<1$.

Finally, to evaluate the CV cluster states in the frequency domain, we consider a squeezing Hamiltonian written in terms of a discrete set of frequency bins, which takes the form
\begin{equation}
    H=i\hbar\kappa \sum_{jk}G_{jk}(a_{i,j}^{\dagger} a_{s,k}^{\dagger} -a_{i,j} a_{s,k}),
\label{eq: Hamiltonian}
\end{equation}
where $a_{\alpha,j}$, $\alpha=i,s$ are the idler (signal) ladder operators of frequency bin $j$, and $G_{jk}$ is called the adjacency matrix of the Hamiltonian graph, connecting modes $j$ and $k$. CV cluster states are defined by a slightly different adjacency matrix, $\mathbf{V}$, also known as the adjacency matrix of the cluster graph. The entries of $\mathbf{V}$ correspond to the weights of the edges of the cluster graph, whereas the modes $\alpha,j$ constitute the vertices. The Hamiltonian and cluster adjacency matrices are mathematically connected \cite{Menicucci2007, Pfister_2019}, such that any Hamiltonian graph $\mathbf{G}$ can define a cluster graph $\mathbf{V}$. For a special case wherein $\mathbf{G}=\mathbf{G}^{-1}$, the two adjacency matrices coincide and $\mathbf{G}=\mathbf{V}$\cite{Menicucci2007, Pfister_2019}.

The signature of CV cluster states is manifested in entanglement witnesses called nullifiers \cite{Menicucci2007,Pfister_2019}. These are linear combinations of the quadrature operators of the different modes $j=1,2,...,N$, given by 
\begin{equation}
q_j (\phi) = e^{i\phi}a_j + e^{-i\phi}a_j^{\dagger},
\label{Qquadrature}
\end{equation}
\begin{equation}
p_j (\phi) = (e^{i\phi}a_j - e^{-i\phi}a_j^{\dagger})/i,
\label{Pquadrature}
\end{equation}
where $\phi$ is a phase factor determining the orientation of the quadrature axis in phase space. Ideally (that is, within the limit of
infinite squeezing) a CV cluster state should satisfy the vector equation 
\begin{equation}
    \braket{(\textbf{p}(\phi)-\mathbf{V}\textbf{q}(\phi))^2} \rightarrow 0,
\label{eq: nul}
\end{equation}
where $\mathbf{q}(\phi)=(q_1,q_2,...,q_N)$ and $\mathbf{p}(\phi)=(p_1,p_2,...,p_N)$ are the quadrature operator vectors. The linear combinations defining the $N$ different nullifiers are the entries of the vector $\mathbf{p}(\phi)-\mathbf{Vq}(\phi)$. Realistically, for finite squeezing it will be necessary to demonstrate simultaneous noise reduction in each of the nullifiers below the vacuum noise level.

\section{Results}
\begin{figure}[h]
\centering
\noindent
\includegraphics[width=0.8\columnwidth]{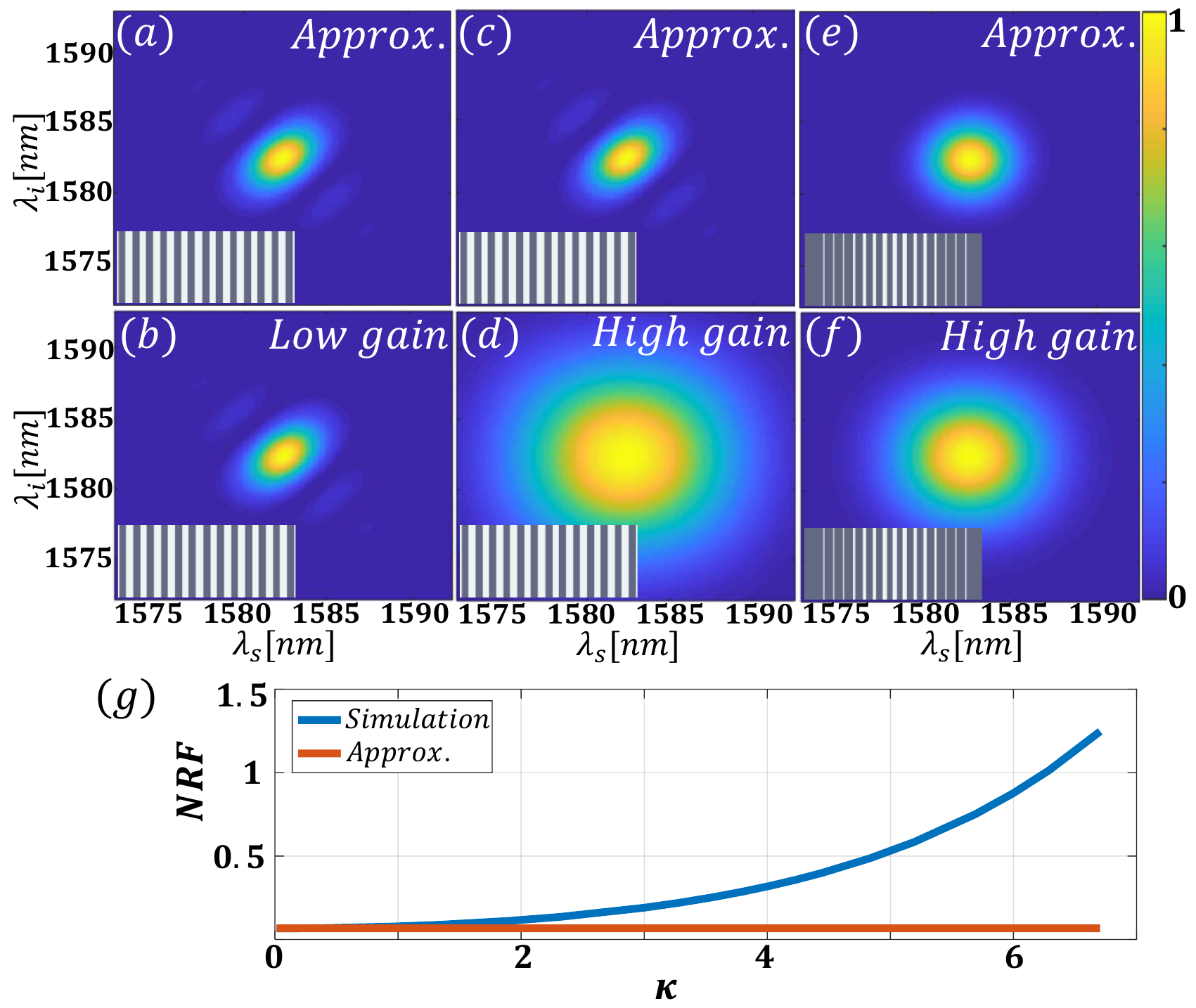}
\caption{Comparison between the JSI and the simulated BSV second-order correlation for 3 different cases. (a-b) low-gain, $\kappa=0.023$ and (c-d) high-gain, $\kappa=22.961$ regimes for a periodically poled KTP crystal; (e-f) round JSI using an apodized crystal in the high-gain regime. The different crystal poling patterns are illustrated in the bottom left insets. (g) NRF as a function of $\kappa$.} 
\label{fig:PPKTP_and_apdized}
\end{figure}

We first validate our model against known results for SPDC spectral shaping in the low-gain regime. In this case, one considers the joint spectral intensity (JSI)\cite{Fedrizzi2020} of the photon pairs, found via first-order perturbation theory as 
\begin{equation}
\mathrm{JSI}=|\mathcal{P}(\omega+\omega') \Phi(\omega,\omega')|^2,
\label{eq: JSI}
\end{equation}
where $\Phi(\omega,\omega')=\int dz \exp{[i\Delta k(\omega,\omega') z]}d_{\mathrm{NL}}(z)$ is the phase-matching function. In the low-gain regime ($\kappa\ll1$), the second-order correlation of BSV given by Eq.\ref{eq: G2_derivation} reduces to the JSI. Fig.\ref{fig:PPKTP_and_apdized}(a-d) shows the JSI alongside our simulation results in the low and high-gain regimes, for a $46\mu m$ periodically poled KTP crystal of length $13.7mm$ pumped by a $791nm$ laser. The BSV correlations in the low-gain regime indeed recover the JSI. 

However, the high-gain regime shows a broader, circularly symmetric correlation, indicating a separable state. This is consistent with the observation that in the high-gain regime the Schmidt rank should decrease \cite{Sharapova2015}. To increase the separability of the squeezed vacuum state throughout all the gain regimes, one may add an apodization to the poling pattern (see insets of Fig.\ref{fig:PPKTP_and_apdized}(e,f)), which prevents side lobes from forming. This results in a circularly-symmetric JSI \cite{Dosseva} that persists also in the high-gain regime (Fig.\ref{fig:PPKTP_and_apdized}(e,f)). 

Since the size of the mode in the frequency domain (the range of frequency over which we measure the photons of each mode) depends on the gain when the latter is increased, it causes a reduction in the number of selected modes captured by the domain of a given window, as was shown in Refs \cite{Shot_Noise_Correlation_High_Gain_PDC,two_color_BSV}. Indeed, when we increase the gain parameter we see a growth of the NRF as shown in Fig.\ref{fig:PPKTP_and_apdized}(g).

\begin{figure}[h!]
\centering
\noindent
\includegraphics[width=\columnwidth]{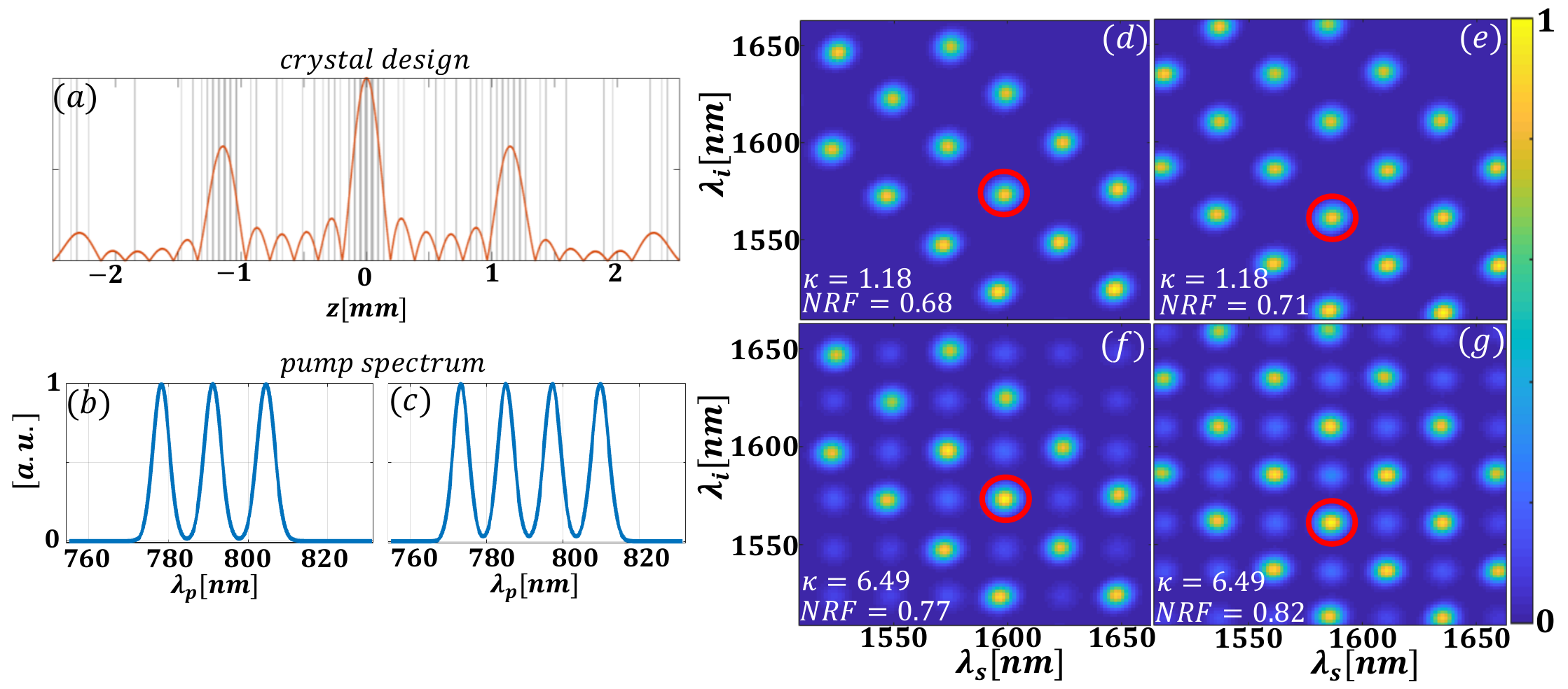}
\caption{High dimensional frequency domain quantum correlations of BSV using ultrafast shaped pump pulses and NLPC's. Different frequency-bin graph states are generated using the same structured NLPC (a), whereas the pump spectrum is varied between three (b) and four (c) Gaussians. (a) Spatial modulation of $\chi^{(2)}$ of the crystal along $z$ (gray), and the Fourier transform of the desired spectrum (brown). Each individual Gaussian has FWHM of $5nm$. Simulated second-order correlation of three (d,f) and four (e,g) Gaussians pump spectrum for different gain parameters, as well as the NRF for the lobes circled in red and the gain parameter.} 
\label{fig:grid}
\end{figure}

Fig.\ref{fig:grid} shows simulated high-dimensional frequency-domain quantum correlations, using the same NLPC hologram (Fig.\ref{fig:grid}(a)) that is pumped with two different spectrally-shaped pumps (Fig.\ref{fig:grid}(b-c)). The two cases shown in Fig.\ref{fig:grid}(d,f) and (e,g) represent, respectively, second-order correlations obtained using a 3-lobe (Fig.\ref{fig:grid}(b)) and 4-lobe (Fig.\ref{fig:grid}(c)) pump spectra, optimized such that each node contains a separable, circularly-symmetric correlation. In the low-gain regime, the shape of these correlations corresponds exactly to the JSI of Eq.\ref{eq: JSI}. For high gain values, additional lobes appear in the second-order correlation function (Fig.\ref{fig:grid}(f,g)), due to the non-negligible probability of higher-order photon pairs. We note that these features do not imply that a decoherence mechanism hampers the quantum state. Rather, they imply that the second order correlations do not encapsulate all the relevant information, as they usually do in the low-gain regime of SPDC. Indeed, a full phase-space description is more suitable for high-gain, and we shall use it to analyze the next example of this section. For completeness, we calculate the NRF for a frequency window containing one node. The NRF in Figs.\ref{fig:grid}(f,g) is higher than in Figs.\ref{fig:grid}(d,e), respectively, as expected due to the increased gain. 

The product of the pump spectrum, $\mathcal{P}(\omega+\omega')$, and the hologram encoded in the phase matching function, $\Phi(\omega,\omega')$,  determines the adjacency matrix $\mathbf{G}$ in the Hamiltonian of Eq.\ref{eq: Hamiltonian}. As explained in the Theory section, the matrix $\mathbf{G}$ defines a Hamiltonian graph state, and, by extension, a cluster state \cite{Pfister_2019}. The ability to arbitrarily engineer and optically control the adjacency matrix $\mathbf{G}$ allows for the modulation of multimode BSV states on an ultrafast time scale, which may prove important for scaling-up CV quantum information processing rates.

\begin{figure}[h]
\centering
\noindent
\includegraphics[width=\columnwidth]{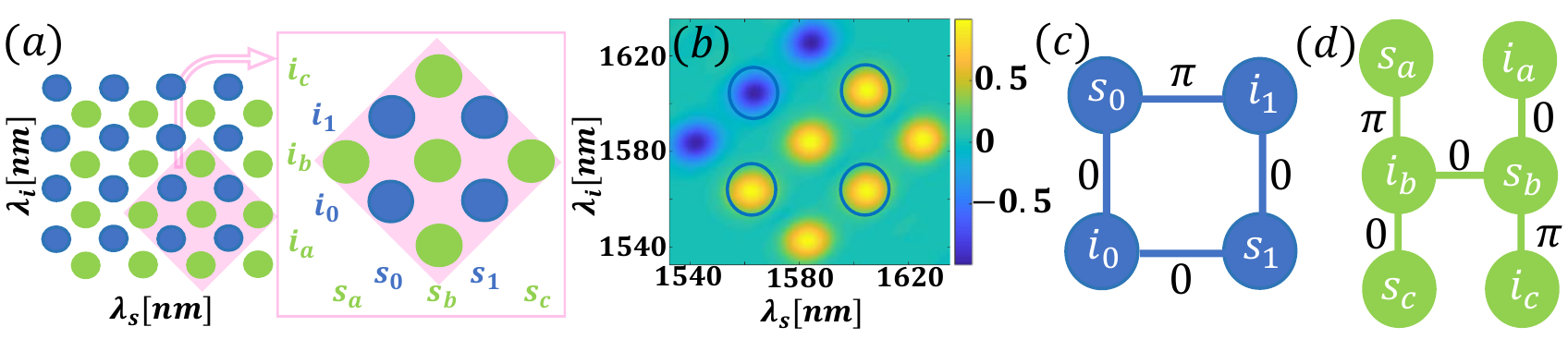}
\caption{(a) Our shaping scheme generates two separable Hamiltonian graph states with square lattice geometry, that are shifted with respect to each other. Spectral filtering or demultiplexing can be used to distinguish between the two graphs. (b) Product of pump spectrum and nonlinear hologram, for the generation of a 4-mode square cluster state (from the highlighted bins) depicted in (c), together with the phase of its weighted edges. The signal and idler each have 2 modes ($s_{0,1}$, $i_{0,1}$), comprising the four nodes of the graph. (d) The graph state that is generated by the second Hamiltonian, together with the phase of its weighted edges. The signal and idler each have 3 modes ($s_{a,b,c}$, $i_{a,b,c}$), comprising the six nodes of the graph.} 
\label{fig:4mode}
\end{figure}

We now focus our attention on the phase-space analysis of such CV cluster states. First, we note that the periodic frequency bin design of Fig.\ref{fig:grid} in fact generates two separable Hamiltonian graph states, with two independent Hamiltonian adjacency matrices $\mathbf{G}^A$ and $\mathbf{G}^B$, such that
\begin{equation}
    H=i\hbar\kappa_A \sum_{jk}G_{jk}^A(a_{i,j}^{\dagger} a_{s,k}^{\dagger} -a_{i,j} a_{s,k})+i\hbar\kappa_B \sum_{jk}G_{jk}^B(b_{i,j}^{\dagger} b_{s,k}^{\dagger} -b_{i,j}b_{s,k})\equiv H_A+H_B.
\label{eq: Hamiltonian square}
\end{equation}
Each of the two Hamiltonian adjacency matrices corresponds to an independent set of modes, forming two non-overlapping square lattices ($A$ and $B$, represented by blue and green circles in Fig.\ref{fig:4mode}(a)) in the idler-signal wavelength plane. Each lattice site corresponds to an \textit{edge} of the graph, connecting a pair of idler-signal modes $a_{i,j},a_{s,k}$ (lattice $A$) or $b_{i,j},b_{s,k}$ (lattice $B$), and each idler or signal mode is a \textit{vertex} in its respective graph. Since these two sets of modes do not overlap, the two parts of the Hamiltonian of Eq.\ref{eq: Hamiltonian square} commute. Therefore, the multimode squeezing operator generating the BSV factors into two separate squeezing operations,
\begin{equation}
S=\exp(-itH/\hbar)=\exp(-itH_A/\hbar)\otimes\exp(-itH_B/\hbar)    
\end{equation}
generating a product of two Hamiltonian graph states. The two Hamiltonian graph states generated by Eq.\ref{eq: Hamiltonian square} are known to be equivalent to cluster states via transformation of the Hamiltonian adjacency matrix into a cluster adjacency matrix \cite{Menicucci2007}. Moreover, their square lattice geometry is known to be universal for MBQC \cite{Pfister_2019}. Each of these cluster states could then be used in parallel for quantum computation. Importantly, as the two clusters form a separable product state, tracing out one cluster state need not hamper the quantum coherence of the other. Below, we focus the discussion on one of the Hamiltonian graph states and analyze the simplest case of an emergent square CV cluster state, first demonstrated in \cite{Menicucci2007}. 

In Fig.\ref{fig:4mode}(b), we plot the product of the pump spectrum and the crystal hologram, which can be used to generate a 4-mode square Hamiltonian graph state in one of the square lattices, marked by blue circles in Fig.\ref{fig:4mode}(b). In this case, the pump spectrum is designed to have three identical equally spaced peaks and the crystal phase matching curve is also designed to have three identical equally spaced peaks. Remarkably, as shown in \cite{Menicucci2007}, in this case, the Hamiltonian graph state is identical to its corresponding cluster state. Fig.\ref{fig:4mode}(c) depicts a graphical representation of the resulting square cluster state. In this example, there are two idler modes, $i_0,i_1$, and two signal modes $s_0,s_1$, at wavelengths $1564nm$ and $1605nm$, respectively, comprising the four vertices of the graph. The values at these frequency bins in Fig.\ref{fig:4mode}(b) correspond to the weighted edges between the four vertices of the graph in Fig.\ref{fig:4mode}(c). For example, vertices $i_1$ and $s_1$ in Fig.\ref{fig:4mode}(c) are connected by an edge with weight $+1/\sqrt{2}$ (phase $0$), the value of which is determined by the value of the marked frequency bin at the top right in Fig.\ref{fig:4mode}(b). Vertices $s_0$ and $i_1$ in Fig.\ref{fig:4mode}(c), however, are connected by an edge with weight $-1/\sqrt{2}$ (phase $\pi$), corresponding to the marked frequency bin at the top left in Fig.\ref{fig:4mode}(b).

For this simple cluster state, denoting $\mathbf{q}(\phi)=(q_{i0}, q_{i1},q_{s0},q_{s1})$ and $\mathbf{p}(\phi)=(p_{i0}, p_{i1}, p_{s0}, p_{s1})$ as the vectors of quadrature operators for the different modes, the Hamiltonian adjacency matrix $\mathbf{G}$ is given by
\begin{equation}
\mathbf{G} = \frac{1}{\sqrt{2}}\begin{pmatrix}
0 & 0 & 1 & 1\\
0 & 0 & -1 & 1\\
1 & -1 & 0 & 0\\
1 & 1 & 0 & 0\\
\end{pmatrix},
\label{eq:adjacency matrix}
\end{equation}
and, as mentioned above, since the Hamiltonian adjacency matrix satisfies the condition $\mathbf{G}=\mathbf{G}^{-1}$, the cluster adjacency matrix, $\mathbf{V}$, is identical to $\mathbf{G}$.

\begin{figure}[h]
\centering
\noindent
\includegraphics[width=\columnwidth]{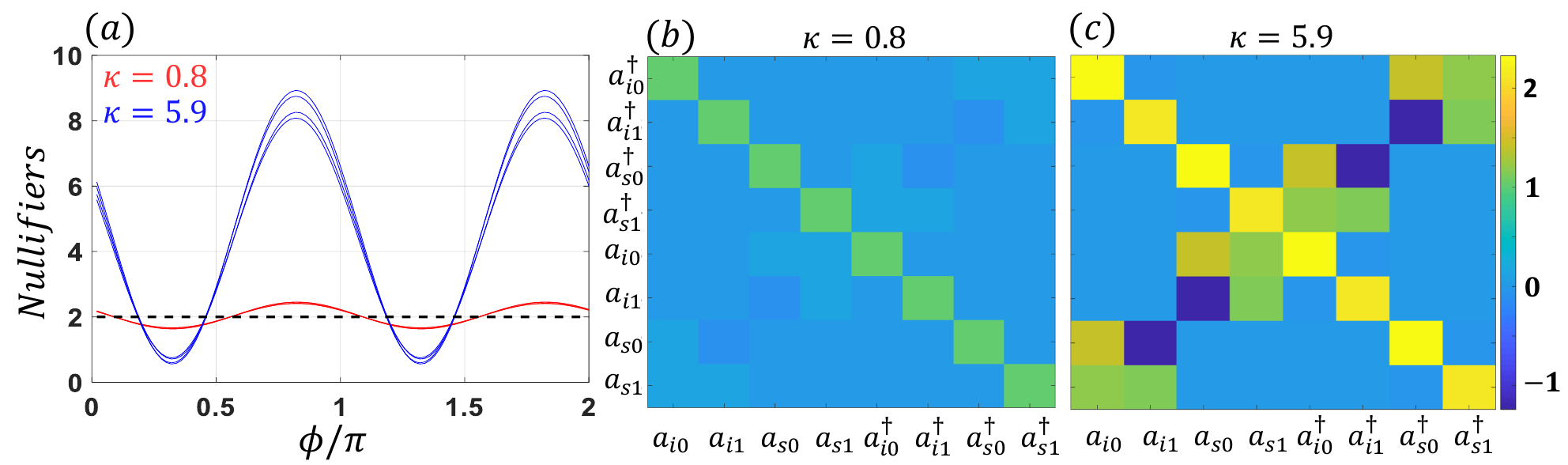}
\caption{(a) Comparison of nullifier uncertainties for two cases of squeezing (red, blue for $\kappa=0.8, 5.9$ respectively). It is evident that for both cases, the nullifier uncertainties fall below the quantum limit (dotted black line) for a specific angle $\phi$. Covariance matrix of the square cluster state for $\kappa=0.8$ (b) and $\kappa=5.9$ (c).} 
\label{fig:covariance}
\end{figure}

To further explore the resulting cluster state, we calculate the uncertainty in each of the four nullifiers, 
\begin{equation}
\begin{split}
\textbf{p}(\phi)-\mathbf{V}\textbf{q}(\phi) = \begin{pmatrix}
 p_{i0}-(q_{s0}+q_{s1})/\sqrt{2} \\ p_{i1}-(q_{s1}-q_{s0})/\sqrt{2} \\ p_{s0}-(q_{i0}-q_{i1})/\sqrt{2} \\ p_{s1}-(q_{i0}+q_{i1})/\sqrt{2}
\end{pmatrix},
\end{split}
\label{eq: nul2}
\end{equation}
which should all be \textit{simultaneously} squeezed below the vacuum noise level to satisfy Eq.\ref{eq: nul}. Fig.\ref{fig:covariance}(a) shows the calculated quantum nullifier uncertainties for the square cluster state, for two different gain parameters $\kappa=0.8, 5.9$. Indeed, in both cases, all four nullifiers are simultaneously squeezed below the vacuum noise level for a certain phase angle $\phi$ of Eqs.\ref{Qquadrature}-\ref{Pquadrature}. Clearly, stronger gain results in a lower nullifier uncertainty, reaching more than $-5dB$ of squeezing. These squeezing levels are already useful for applications in sensing and spectroscopy.


For completeness, we calculate the covariance matrix (Eq.\ref{eq:covariance}) determining the Wigner function (Eq.\ref{eq: Wigner}) of our square cluster state (Fig.\ref{fig:4mode}(c)), as shown in Fig.\ref{fig:covariance}(b,c) for two different gain parameters. For low gain, the squeezing is weak and the covariance matrix is expected to resemble the identity matrix - the covariance matrix of the vacuum state\cite{Adesso_2014}. As the gain increases, additional non-zero off-diagonal elements appear in the covariance matrix, indicating correlations between states in the chosen basis. Note that, for similar reasons used to explain Eq.\ref{eq: G1 fields} and Eq.\ref{eq: Q_ij}, all elements of the form $\braket{a_{i,j}a_{i,k}},\braket{a_{s,j} a_{s,k}},\braket{a_{i,j}^\dagger a_{s,k}}$ for $j,k=0,1$ and their conjugates, are identically zero.

\section{Conclusions}
In summary, we proposed design methods for spectral shaping of BSV using tailored NLPC holograms and ultrashort, structured, and intense pump beams, for applications in frequency-domain continuous variable quantum information processing. We provide a series of tools to characterize the generated states, including the signal-idler first and second-order correlations (Eqs.\ref{eq: G1 fields},\ref{eq: G2_derivation}), the noise reduction factor (Eq.\ref{eq: NRF}), the covariance matrix (Eq.\ref{eq:covariance}) and the nullifier uncertainties (Eq.\ref{eq: nul}).

Contrary to the conventional methods for generating frequency domain CV cluster states employing the frequency comb of optical parametric oscillators, our proposed nonlinear holograms \cite{Shiloh}, which could also be accompanied with quasiperiodic modulation \cite{Lifshitz_Arie_Bahabad}, enable to arbitrarily control the phase matching function in the frequency domain. Further, the all-optical control allowed by our scheme enables the modulation of such CV cluster states on ultrafast times scales. Our method can serve as a design tool for quantum information applications that utilize the spectral degree of freedom of light. Owing to the high degree of control over the output state - by the pump spectrum and crystal holograms - inverse design optimization of modular quantum correlations over frequency-bin grids can be readily employed to discover further designs of desired quantum states, as recently done for the spatial degree of freedom of SPDC \cite{Rozenberg,Kovlakov_PRA_2018}. 

Moreover, our proposed method can also be extended to generate energy-time CV cluster states by exploiting the interplay between the spectral and temporal degrees of freedom of light. This can be achieved by incorporating tailored temporal modulation of the pump beam with the NLPC, which would enable us to engineer the joint spectral-temporal amplitude of the generated BSV. By combining our proposed method with low-frequency temporal modulation techniques \cite{Furusawa}, we can generate a wide variety of high-dimensional CV cluster states for various applications in quantum information processing.

\begin{backmatter}
\bmsection{Funding}
Ministry of Science, Technology and Space; Israel Science Foundation (969/22).

\bmsection{Acknowledgments}
A.K. acknowledges support by the Adams Fellowship of the Israeli Academy of Sciences and Humanities.

\bmsection{Disclosures}
The authors declare no conflicts of interest.

\bmsection{Data Availability Statement}
Data underlying the results presented in this paper are not publicly available at this time but may be obtained from the authors upon reasonable request.


\end{backmatter}



\bibliography{references}

\begin{thebibliography}{10}
\newcommand{\enquote}[1]{``#1''}

\bibitem{Clemmen2016}
S.~Clemmen, A.~Farsi, S.~Ramelow, and A.~L. Gaeta, \enquote{Ramsey interference
  with single photons,} {\protect\JournalTitle{Phys. Rev. Lett.}} \textbf{117},
  223601 (2016).

\bibitem{Karnieli2018}
A.~Karnieli and A.~Arie, \enquote{Frequency domain stern gerlach effect for
  photonic qubits and qutrits,} {\protect\JournalTitle{Optica}} \textbf{5},
  1297--1303 (2018).

\bibitem{Shaked2018}
Y.~Shaked, Y.~Michael, R.~Z. Vered, L.~Bello, M.~Rosenbluh, and A.~Pe'er,
  \enquote{{Lifting the bandwidth limit of optical homodyne measurement with
  broadband parametric amplification},} {\protect\JournalTitle{Nature
  Communications}} \textbf{9}, 609 (2018).

\bibitem{Bloch2007}
M.~Bloch, S.~W. McLaughlin, J.~M. Merolla, and F.~Patois,
  \enquote{Frequency-coded quantum key distribution,}
  {\protect\JournalTitle{Opt. Lett.}} \textbf{32}, 301--303 (2007).

\bibitem{Kues2019}
M.~Kues, C.~Reimer, J.~M. Lukens, W.~J. Munro, A.~M. Weiner, D.~J. Moss, and
  R.~Morandotti, \enquote{{Quantum optical microcombs},}
  {\protect\JournalTitle{Nature Photonics}} \textbf{13}, 170--179 (2019).

\bibitem{Lu2022}
H.~H. Lu, K.~V. Myilswamy, R.~S. Bennink, S.~Seshadri, M.~S. Alshaykh, J.~Liu,
  T.~J. Kippenberg, D.~E. Leaird, A.~M. Weiner, and J.~M. Lukens,
  \enquote{{Bayesian tomography of high-dimensional on-chip biphoton frequency
  combs with randomized measurements},} {\protect\JournalTitle{Nature
  Communications}} \textbf{13} (2022).

\bibitem{Kues2017}
M.~Kues, C.~Reimer, P.~Roztocki, L.~R. Cort{\'{e}}s, S.~Sciara, B.~Wetzel,
  Y.~Zhang, A.~Cino, S.~T. Chu, B.~E. Little, D.~J. Moss, L.~Caspani,
  J.~Aza{\~{n}}a, and R.~Morandotti, \enquote{{On-chip generation of
  high-dimensional entangled quantum states and their coherent control},}
  {\protect\JournalTitle{Nature}} \textbf{546}, 622--626 (2017).

\bibitem{entanglement_of_60_modes_Pfister}
M.~Chen, N.~C. Menicucci, and O.~Pfister, \enquote{Experimental realization of
  multipartite entanglement of 60 modes of a quantum optical frequency comb,}
  {\protect\JournalTitle{Phys. Rev. Lett.}} \textbf{112}, 120505 (2014).

\bibitem{Reimer2019}
C.~Reimer, S.~Sciara, P.~Roztocki, M.~Islam, L.~{Romero Cort{\'{e}}s},
  Y.~Zhang, B.~Fischer, S.~Loranger, R.~Kashyap, A.~Cino, S.~T. Chu, B.~E.
  Little, D.~J. Moss, L.~Caspani, W.~J. Munro, J.~Aza{\~{n}}a, M.~Kues, and
  R.~Morandotti, \enquote{{High-dimensional one-way quantum processing
  implemented on d-level cluster states},} {\protect\JournalTitle{Nature
  Physics}} \textbf{15}, 148--153 (2019).

\bibitem{Chekhova_Science_2022}
T.~Santiago-Cruz, S.~D. Gennaro, O.~Mitrofanov, S.~Addamane, J.~Reno,
  I.~Brener, and M.~V. Chekhova, \enquote{Resonant metasurfaces for generating
  complex quantum states,} {\protect\JournalTitle{Science}} \textbf{377},
  991--995 (2022).

\bibitem{Dosseva}
A.~Dosseva, L.~Cincio, and A.~M. Bra\ifmmode~\acute{n}\else \'{n}\fi{}czyk,
  \enquote{Shaping the joint spectrum of down-converted photons through
  optimized custom poling,} {\protect\JournalTitle{Phys. Rev. A}} \textbf{93},
  013801 (2016).

\bibitem{Graffitti2017}
F.~Graffitti, D.~Kundys, D.~T. Reid, A.~M. Bra{\'{n}}czyk, and A.~Fedrizzi,
  \enquote{{Pure down-conversion photons through sub-coherence-length domain
  engineering},} {\protect\JournalTitle{Quantum Science and Technology}}
  \textbf{2} (2017).

\bibitem{Helt2015}
L.~G. Helt, M.~J. Steel, and J.~E. Sipe, \enquote{{Spontaneous parametric
  downconversion in waveguides: What's loss got to do with it?}}
  {\protect\JournalTitle{New Journal of Physics}} \textbf{17} (2015).

\bibitem{SFWM_2023}
K.~Garay-Palmett, D.~B. Kim, Y.~Zhang, F.~A. Dom\'{i}nguez-Serna, V.~O. Lorenz,
  and A.~B. U'Ren, \enquote{Fiber-based photon-pair generation: tutorial,}
  {\protect\JournalTitle{J. Opt. Soc. Am. B}} \textbf{40}, 469--490 (2023).

\bibitem{Da_Zhang}
D.~Zhang, C.~Li, Z.~Zhang, Y.~Zhang, Y.~Zhang, and M.~Xiao, \enquote{Enhanced
  intensity-difference squeezing via energy-level modulations in hot atomic
  media,} {\protect\JournalTitle{Phys. Rev. A}} \textbf{96}, 043847 (2017).

\bibitem{Shiloh}
R.~Shiloh and A.~Arie, \enquote{Spectral and temporal holograms with nonlinear
  optics,} {\protect\JournalTitle{Opt. Lett.}} \textbf{37}, 3591--3593 (2012).

\bibitem{Fedrizzi2022}
C.~L. Morrison, F.~Graffitti, P.~Barrow, A.~Pickston, J.~Ho, and A.~Fedrizzi,
  \enquote{Frequency-bin entanglement from domain-engineered down-conversion,}
  {\protect\JournalTitle{APL Photonics}} \textbf{7}, 066102 (2022).

\bibitem{Graffitti:18}
F.~Graffitti, P.~Barrow, M.~Proietti, D.~Kundys, and A.~Fedrizzi,
  \enquote{Independent high-purity photons created in domain-engineered
  crystals,} {\protect\JournalTitle{Optica}} \textbf{5}, 514--517 (2018).

\bibitem{Sharapova2018}
P.~R. Sharapova, O.~V. Tikhonova, S.~Lemieux, R.~W. Boyd, and M.~V. Chekhova,
  \enquote{{Bright squeezed vacuum in a nonlinear interferometer: Frequency and
  temporal Schmidt-mode description},} {\protect\JournalTitle{Physical Review
  A}} \textbf{97}, 053827 (2018).

\bibitem{Pfister_2019}
O.~Pfister, \enquote{Continuous-variable quantum computing in the quantum
  optical frequency comb,} {\protect\JournalTitle{Journal of Physics B: Atomic,
  Molecular and Optical Physics}} \textbf{53}, 012001 (2019).

\bibitem{Supersensitive_Phase_Meas2017}
M.~Manceau, G.~Leuchs, F.~Khalili, and M.~Chekhova, \enquote{Detection loss
  tolerant supersensitive phase measurement with an su(1,1) interferometer,}
  {\protect\JournalTitle{Phys. Rev. Lett.}} \textbf{119}, 223604 (2017).

\bibitem{Menicucci2007}
N.~C. Menicucci, S.~T. Flammia, H.~Zaidi, and O.~Pfister, \enquote{Ultracompact
  generation of continuous-variable cluster states,}
  {\protect\JournalTitle{Phys. Rev. A}} \textbf{76}, 010302 (2007).

\bibitem{Cai2017}
Y.~Cai, J.~Roslund, G.~Ferrini, F.~Arzani, X.~Xu, C.~Fabre, and N.~Treps,
  \enquote{{Multimode entanglement in reconfigurable graph states using optical
  frequency combs},} {\protect\JournalTitle{Nature Communications}} \textbf{8},
  1--9 (2017).

\bibitem{Drago2022}
C.~Drago and A.~M. Bra\ifmmode~\acute{n}\else \'{n}\fi{}czyk, \enquote{Tunable
  frequency-bin multimode squeezed vacuum states of light,}
  {\protect\JournalTitle{Phys. Rev. A}} \textbf{106}, 043714 (2022).

\bibitem{Treps2018}
F.~Arzani, C.~Fabre, and N.~Treps, \enquote{Versatile engineering of multimode
  squeezed states by optimizing the pump spectral profile in spontaneous
  parametric down-conversion,} {\protect\JournalTitle{Phys. Rev. A}}
  \textbf{97}, 033808 (2018).

\bibitem{Nokkala_2018}
J.~Nokkala, F.~Arzani, F.~Galve, R.~Zambrini, S.~Maniscalco, J.~Piilo,
  N.~Treps, and V.~Parigi, \enquote{Reconfigurable optical implementation of
  quantum complex networks,} {\protect\JournalTitle{New Journal of Physics}}
  \textbf{20}, 053024 (2018).

\bibitem{Drummond1995}
M.~J. Werner, M.~G. Raymer, M.~Beck, and P.~D. Drummond, \enquote{Ultrashort
  pulsed squeezing by optical parametric amplification,}
  {\protect\JournalTitle{Phys. Rev. A}} \textbf{52}, 4202--4213 (1995).

\bibitem{Brambilla}
E.~Brambilla, A.~Gatti, M.~Bache, and L.~A. Lugiato, \enquote{Simultaneous
  near-field and far-field spatial quantum correlations in the high-gain regime
  of parametric down-conversion,} {\protect\JournalTitle{Phys. Rev. A}}
  \textbf{69}, 023802 (2004).

\bibitem{Sharapova2020}
P.~R. Sharapova, G.~Frascella, M.~Riabinin, A.~M. P\'erez, O.~V. Tikhonova,
  S.~Lemieux, R.~W. Boyd, G.~Leuchs, and M.~V. Chekhova, \enquote{Properties of
  bright squeezed vacuum at increasing brightness,}
  {\protect\JournalTitle{Phys. Rev. Research}} \textbf{2}, 013371 (2020).

\bibitem{TM2020}
S.~Trajtenberg-Mills, A.~Karnieli, N.~Voloch-Bloch, E.~Megidish, H.~S.
  Eisenberg, and A.~Arie, \enquote{Simulating correlations of structured
  spontaneously down-converted photon pairs,} {\protect\JournalTitle{Laser \&
  Photonics Reviews}} \textbf{14}, 1900321 (2020).

\bibitem{Rozenberg}
E.~Rozenberg, A.~Karnieli, O.~Yesharim, J.~Foley-Comer, S.~Trajtenberg-Mills,
  D.~Freedman, A.~M. Bronstein, and A.~Arie, \enquote{Inverse design of
  spontaneous parametric downconversion for generation of high-dimensional
  qudits,} {\protect\JournalTitle{Optica}} \textbf{9}, 602--615 (2022).

\bibitem{Garrison2008}
J.~C. Garrison and R.~Y. Chiao, \emph{{Quantum Optics}} (Oxford University,
  2008).

\bibitem{Adesso_2014}
G.~Adesso, S.~Ragy, and A.~R. Lee, \enquote{Continuous variable quantum
  information: Gaussian states and beyond,} {\protect\JournalTitle{arXiv}}
  \textbf{21}, 1440001 (2014).

\bibitem{Sipe2014}
N.~Quesada and J.~E. Sipe, \enquote{Effects of time ordering in quantum
  nonlinear optics,} {\protect\JournalTitle{Phys. Rev. A}} \textbf{90}, 063840
  (2014).

\bibitem{Sipe2020}
N.~Quesada, G.~Triginer, M.~D. Vidrighin, and J.~E. Sipe, \enquote{Theory of
  high-gain twin-beam generation in waveguides: From maxwell's equations to
  efficient simulation,} {\protect\JournalTitle{Phys. Rev. A}} \textbf{102},
  033519 (2020).

\bibitem{Christ2013}
A.~Christ, B.~Brecht, W.~Mauerer, and C.~Silberhorn, \enquote{{Theory of
  quantum frequency conversion and type-II parametric down-conversion in the
  high-gain regime},} {\protect\JournalTitle{New Journal of Physics}}
  \textbf{15} (2013).

\bibitem{two_color_BSV}
I.~N. Agafonov, M.~V. Chekhova, and G.~Leuchs, \enquote{Two-color bright
  squeezed vacuum,} {\protect\JournalTitle{Phys. Rev. A}} \textbf{82}, 011801
  (2010).

\bibitem{Fedrizzi2020}
F.~Graffitti, P.~Barrow, A.~Pickston, A.~M. Bra\ifmmode~\acute{n}\else
  \'{n}\fi{}czyk, and A.~Fedrizzi, \enquote{Direct generation of tailored
  pulse-mode entanglement,} {\protect\JournalTitle{Phys. Rev. Lett.}}
  \textbf{124}, 053603 (2020).

\bibitem{Sharapova2015}
P.~Sharapova, A.~M. P\'erez, O.~V. Tikhonova, and M.~V. Chekhova,
  \enquote{Schmidt modes in the angular spectrum of bright squeezed vacuum,}
  {\protect\JournalTitle{Phys. Rev. A}} \textbf{91}, 043816 (2015).

\bibitem{Shot_Noise_Correlation_High_Gain_PDC}
O.~Jedrkiewicz, Y.-K. Jiang, E.~Brambilla, A.~Gatti, M.~Bache, L.~A. Lugiato,
  and P.~Di~Trapani, \enquote{Detection of sub-shot-noise spatial correlation
  in high-gain parametric down conversion,} {\protect\JournalTitle{Phys. Rev.
  Lett.}} \textbf{93}, 243601 (2004).

\bibitem{Lifshitz_Arie_Bahabad}
R.~Lifshitz, A.~Arie, and A.~Bahabad, \enquote{Photonic quasicrystals for
  nonlinear optical frequency conversion,} {\protect\JournalTitle{Phys. Rev.
  Lett.}} \textbf{95}, 133901 (2005).

\bibitem{Kovlakov_PRA_2018}
E.~V. Kovlakov, S.~S. Straupe, and S.~P. Kulik, \enquote{Quantum state
  engineering with twisted photons via adaptive shaping of the pump beam,}
  {\protect\JournalTitle{Phys. Rev. A}} \textbf{98}, 060301 (2018).

\bibitem{Furusawa}
J.~Yoshikawa, S.~Yokoyama, T.~Kaji, C.~Sornphiphatphong, Y.~Shiozawa,
  K.~Makino, and A.~Furusawa, \enquote{Invited article: Generation of
  one-million-mode continuous-variable cluster state by unlimited time-domain
  multiplexing,} {\protect\JournalTitle{APL Photonics}} \textbf{1}, 060801
  (2016).

\end{thebibliography}





\end{document}